\documentclass{article}
\usepackage{spconf,amsmath,graphicx}

\usepackage{hyperref}
\usepackage[utf8]{inputenc}
\usepackage{tabularx}
\usepackage{setspace}
\usepackage{graphicx}
\usepackage{booktabs}
\usepackage{amsmath}
\usepackage{color}
\usepackage{multirow}

\usepackage{url,cite} 
\usepackage{caption}
\usepackage{verbatim} 
\usepackage{amssymb}
\usepackage{algorithm}
\usepackage{algorithmic}
\usepackage{tikz}
\usepackage{footnote}
\usepackage{url}

\title{SVSGAN: Singing Voice Separation via Generative Adversarial Network}
\name{Zhe-Cheng Fan, Yen-Lin Lai, Jyh-Shing R. Jang}
\address{Department of Computer Science and Information Engineering, National Taiwan University, Taiwan}

\begin{document}

%
\maketitle
\begin{abstract}

Separating two sources from an audio mixture is an important task with many applications. It is a challenging problem since only one signal channel is available for analysis. In this paper, we propose a novel framework for singing voice separation  using the generative adversarial network (GAN) with a time-frequency masking function. The mixture spectra is considered to be a distribution and is mapped to the clean spectra which is also considered a distribtution. The approximation of distributions between mixture spectra and clean spectra is performed during the adversarial training process. 
In contrast with current deep learning approaches for source separation, the parameters of the proposed framework are first initialized in a supervised setting and then optimized by the training procedure of GAN in an unsupervised setting. Experimental results on three datasets (MIR-1K, iKala and DSD100) show that performance can be improved by the proposed framework consisting of conventional networks.

\end{abstract}
\section{Introduction}
\label{sec:intro}

Monaural source separation is important to various music applications and is sometimes used as a pre-processing step of music signal analysis. For instance, leading instrument detection \cite{durrieu11jstsp, stefan15icassp} separates a leading instrument from its accompaniments. Singing pitch estimation \cite{hsu12taslp,fan16bigmm,yoshii16taslp} can be improved by first separating vocals from background music. Cover song identification \cite{serra10amir} is also based on leading instrument or vocal pitch features, estimated using a separated singing voice. 

\begin{figure}[t]
\begin{center}
\includegraphics[trim=6cm 4cm 8cm 3cm,clip,scale=0.4]{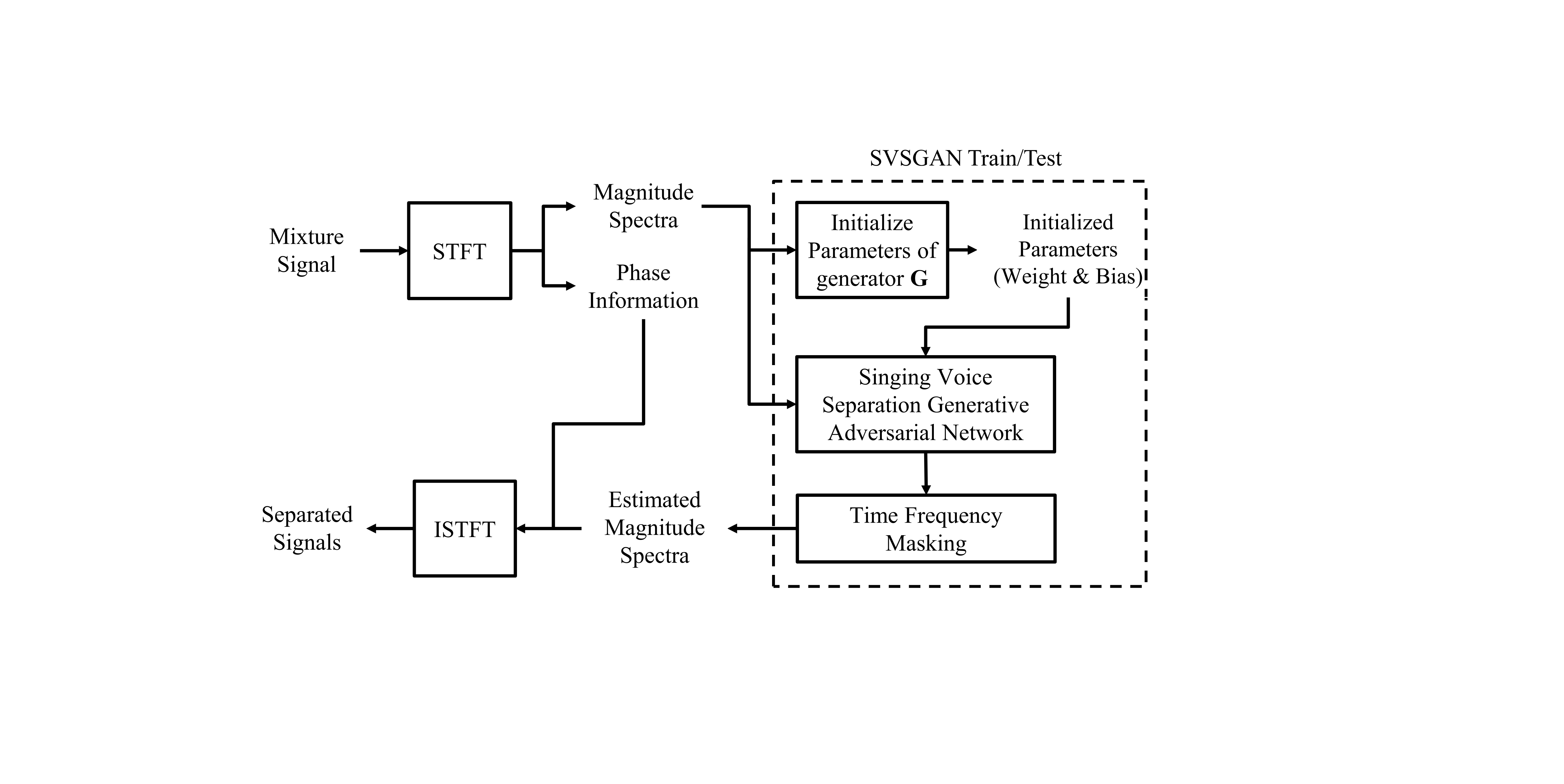}
\caption{Block diagram of the proposed framework.}
\label{fig:block diagram}
\vspace{-10pt}
\end{center}
\end{figure}

Several approaches have been proposed for singing voice separation. Rafii and Pardo proposed the REPET system \cite{raffii13taslp} to separate voice and music by extracting the repeating musical structure. Assumption of low rank and sparsity of music has been used for matrix decompostition\cite{huang12icassp, sprechmann12ismir, yang13ismir, chan15icassp}. The widely used non-negative matrix factorization (NMF) is applied by learning the non-negative reconstruction bases and weights for singing voice separation \cite{zhu13taslp}. Moreover, a complex NMF model \cite{magron16icassp} has been proposed for jointly estimating the spectrogram and the source phase.

With the development of deep learning, Mass \emph{et al.} \cite{mass12interspeech} used recurrent neural networks (RNN) to create a clean voice. Huang \emph{et al.} \cite{huang14ismir} then proposed deep RNN with discriminative training to reconstruct vocals from background music. Training multi-context networks \cite{zhang15icassp, zhang16taslp} with different inputs combined at layer level was proposed to improve audio separation performance. Deep clustering \cite{luo17icassp} is also used for music separation. Post-processing with a Wiener filter at the output of neural networks and data augmentation \cite{stefan17icassp} have been proposed to separate vocals and instruments. All of these deep learning techniques use multiple non-linear layers to learn the optimal hidden representations from data in a supervised setting. 

\begin{figure*}[t]
\begin{center}
\vspace{-20pt}
\includegraphics[trim=0.3cm 2.4cm 1cm 2cm,clip,scale=0.45]{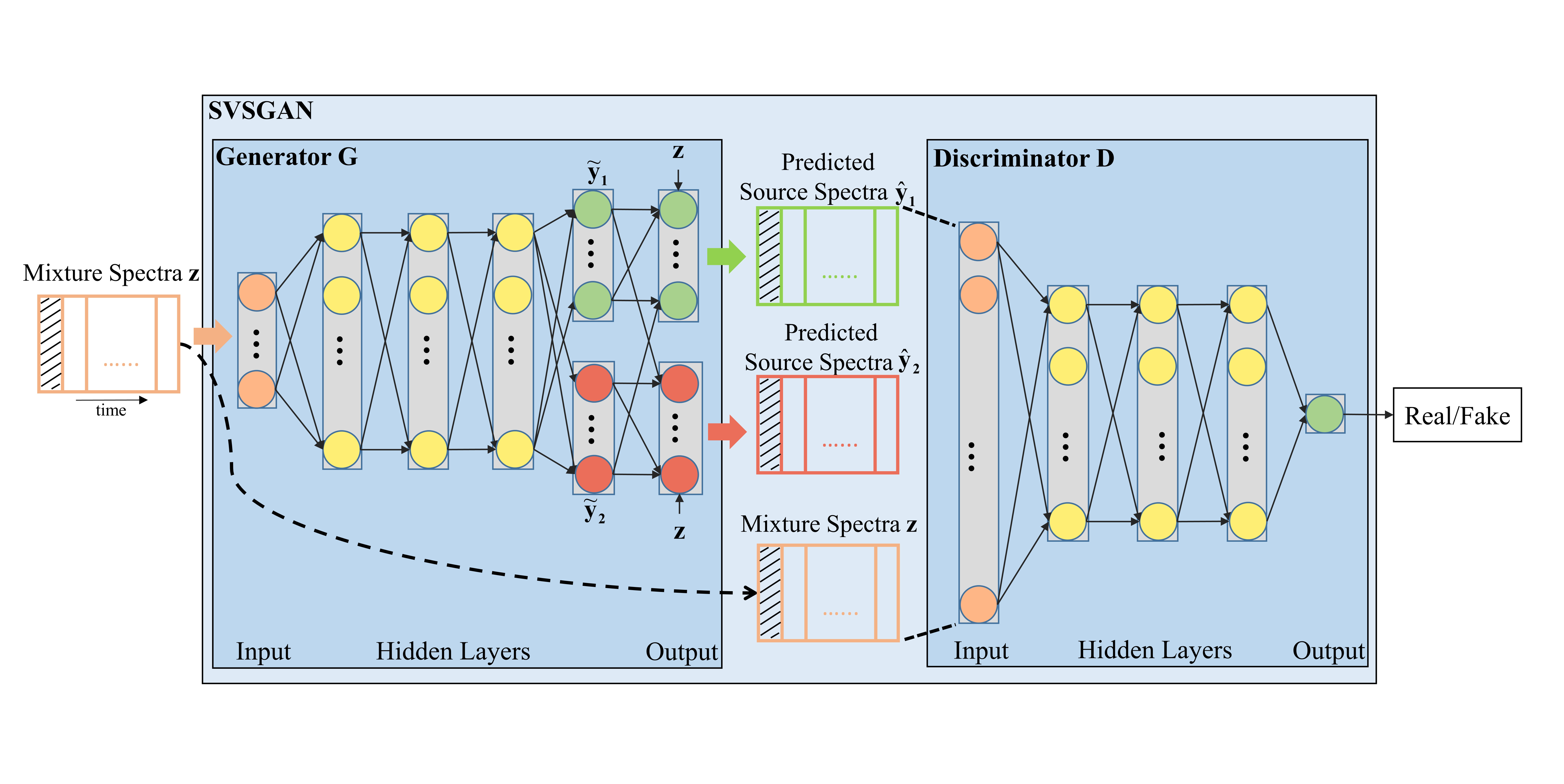}
\caption{The proposed SVSGAN framework which consists of two conventional DNNs: generator G and discriminator D. Each spectrum is considered to be a sample vector coming from a distribution of spectra.}
\label{fig:svsgan block diagram}
\vspace{-10pt}
\end{center}
\end{figure*}

Generative adversarial networks (GAN) are a new generative model of deep learning \cite{goodfellow14nips}, which has been successfully used in the field of computer vision to generate realistic images. In the field of source separation, Pascual \emph{et al.} \cite{serra17interspeech} proposed the use of GAN on speech enhancement, which operates in the waveform domain and aims to generate clean vocal waveforms. This paper proposes a novel framework for singing voice separation via GAN (SVSGAN) which operates in the frequency domain and uses a conditional version of GAN. To our knowledge, this is the first proposed framework to use adversarial learing to perform sinigng voice separation.
We regard each spectrum as a sample vector coming from the distribution of spectra. Non-linear mapping of distributions between the mixture spectra and the clean spectra is performed during the adversarial training process. 
Before adversarial training, the generator parameters are first initialized with joint optimization in a supervised setting and then optimized by the SVSGAN training process in an unsupervised setting. Finally the time-frequency masking function consists of the generator outputs. The block diagram of the proposed framework is shown in Fig.~\ref{fig:block diagram}.

The remainder of this paper is organized as follows: Section \ref{sec:gan} gives an overview of GAN. Section \ref{sec:svsgan} presents the deatails of proposed model including parameter initialization and the adversarial training process. Section \ref{sec:experiments} presents the experimental settings and results using the MIR-1K, iKala and DSD100 datasets. We conclude the paper in Section \ref{sec:concl&future work}.

\section{Generative Adversarial Networks}
\label{sec:gan}

Ian \emph{et al.} \cite{goodfellow14nips} proposed adversarial learning models that learn to map samples $z$ from one distribution to samples $x$ from another distribution. GAN consists of generative model G and discriminative model D, which compete in a two-player min-max game. G aims to imitate the real data distribution while D is a binary classifier which tries to accurately distinguish real data from those generated. Within this min-max game, the generator and the discriminator can be trained jointly by optimizing the following objective function:
\begin{equation}
\begin{split}
\min_G\max_D V(G,D)=E_{x \sim P_{data(x)}}[\log D(x)] + \\
 E_{z \sim P_{G(z)}}[\log (1-D(G(z)))], 
\end{split}
\end{equation}
where $x$ is real data sampled from distribution $P_{data}$ and $G(z)$ stands for artificial data sampled from distribution $P_{G}$. It is shown in \cite{goodfellow14nips} that sufficient training data and epochs allows the distribution $P_{G}$ to coverage to the distribution $P_{data}$. 

To get more mapping information, we use a conditional extension of GAN (CGAN) \cite{isola16arxiv} which is augmented with some side information. Suppose there is a context vector $y$ as side information, the generator $G(z, y)$ tries to synthesize realistic data under the control of $y$. Similarly, the CGAN model allows the output of the discriminative model $D(x, y)$ to be controlled by the context vector $y$. The objective function becomes the following: 
\begin{equation}
\label{eq:cgan}
\begin{split}
\min_G\max_D V_{CGAN}(G,D)=E_{x,y \sim P_{data(x,y)}}[\log D(x,y)] + \\
 E_{z \sim P_{G(z)}, y \sim P_{data}(y)}[\log (1-D(G(z,y),y))].
\end{split}
\end{equation}
In this work, we adjust the input of CGAN, which is discussed in Section \ref{sec:svsgan}.

\section{Proposed Work}
\label{sec:svsgan}

\subsection{Model of Singing Voice Separation GAN (SVSGAN)}

The SVSGAN architecture consists of two conventional deep neural networks (DNNs): generator G and discriminator D, as shown in Fig.~\ref{fig:svsgan block diagram}. We use magnitude spectra as features and take each spectrum as a sample vector from the spectra distribution. Non-linear mapping is performed between the input mixture spectrum and output clean spectrum, which consists of the vocal part and background music part. Generator G inputs a mixture spectra and generates realistic vocal and background music spectra while discriminator D distinguishes the clean spectra from those generated spectra.

Given that magnitude spectra are transformed from the time domain audio signals using short time Fourier transform (STFT), the output targets $\mathbf{y_1}$ and $\mathbf{y_2}$ of the network are the magnitude spectra of different sources. After training, the network's output predictions, which are also magnitude spectra, are $\mathbf{\tilde{y}_1}$ and $\mathbf{\tilde{y}_2}$. The time-frequency masking function, called a soft time-frequency mask, can smooth the source separation results and is used here. The time-frequency mask can be defined as:  
\begin{equation}
\mathbf{m}(f)=\frac{\vert\mathbf{\tilde{y}_1}(f)\vert}{\vert\mathbf{\tilde{y}_1}(f)\vert+\vert\mathbf{\tilde{y}_2}(f)\vert},
\end{equation}
where $f=1,2, ..., \mathbf{\textit{F}}$, stands for different frequencies. After a time-frequency mask is calculated, it is applied to the spectra $\mathbf{z}$ of the mixture signals to estimate the predicted separation spectra $\mathbf{\tilde{s}_1}$ and $\mathbf{\tilde{s}_2}$, corresponding to source $1$ and source $2$, defined as:
\begin{equation}
\begin{split} 
\label{eq:time-frequency mask}
&\mathbf{\tilde{s}_1}(f)=\mathbf{m}(f)\mathbf{z}(f), \\
&\mathbf{\tilde{s}_2}(f)=(1-\mathbf{m}(f))\mathbf{z}(f),
\end{split}
\end{equation}
where $f=1,2, ..., \mathbf{\textit{F}}$, stands for different frequencies. 
However, based on \cite{huang14ismir}, the joint optimization is proposed to achieve better results. Similarly, instead of training the network for the time-frequency mask, we train it with the time-frequency masking function. As shown in the left part of Fig.\ref{fig:svsgan block diagram}, the time-frequency masking function is regarded as an additional layer at the network output,  defined as:
\begin{equation}
\begin{split} 
\label{eq:joint-training}
&\mathbf{\hat{y}_1}=\frac{\vert\mathbf{\tilde{y}_1}\vert}{\vert\mathbf{\tilde{y}_1}\vert+\vert\mathbf{\tilde{y}_2}\vert}\otimes\mathbf{z}, \\
&\mathbf{\hat{y}_2}=\frac{\vert\mathbf{\tilde{y}_2}\vert}{\vert\mathbf{\tilde{y}_1}\vert+\vert\mathbf{\tilde{y}_2}\vert}\otimes\mathbf{z}, 
\end{split}
\end{equation}
where $\otimes$ stands for elementwise operation. $\mathbf{\hat{y}_1}$ and $\mathbf{\hat{y}_2}$ are estimated spectra, which can be transformed into time-domain signals using the inverse short time Fourier transform (ISTFT) with phase information. In this way, the network and time-frequency masking function are jointly optimized. In our proposed framework, the final output separated spectra is based on Eq.~\ref{eq:joint-training}.

\subsection{Training Objective Functions}

Before adversarial training, 
the parameters of the generator G are initialized by performing Eq.~\ref{eq:joint-training} in a supervised setting. The training obejctive $J$ is the mean squared error (MSE) function, which is defined as follows:
\begin{equation}
\label{eq:J}
J=\Arrowvert \mathbf{\hat{y}_1}-\mathbf{y}_1 \Arrowvert^2+\Arrowvert \mathbf{\hat{y}_2}-\mathbf{y}_2 \Arrowvert^2.
\end{equation}
After parameter initialization, the generator G provides basic performance for singing voice separation to serve as the baseline for our experiments.

To fit the input of generator G, the training objective function of SVSGAN by adjusting Eq.~\ref{eq:cgan} is defined as follows:
\begin{equation}
\begin{split}
&\min_G\max_D V_{SVSGAN}(G,D)=\\
&E_{\mathbf{z},\mathbf{s}_c \sim P_{data(\mathbf{z},\mathbf{s}_c)}}[\log D(\mathbf{s}_c, \mathbf{z})] + \\
&E_{\mathbf{z} \sim P_{G(\mathbf{z})}}[\log (1-D(G(\mathbf{z}), \mathbf{z})], 
\end{split}
\end{equation}
where $\mathbf{s}_c$ is the concatenation of $\mathbf{y_1}$ and $\mathbf{y_2}$, and the output of $G(\mathbf{z})$ is the predicted spectra consisting of the concatenation of $\mathbf{\hat{y}_1}$ and $\mathbf{\hat{y}_2}$, which is generated from input spectra $\mathbf{z}$. The output of discirminator D is controlled by the augmented input spectra $\mathbf{z}$. By this step, the SVSGAN not only approximates the distribution between input spectra and output spectra but also learns the general structure of the spectra. In addition, we use log $D$ trick\cite{goodfellow14nips} as the objective function for generator G.

Note that better separation results may be obtained using complicated training objective functions and more powerful neural networks, such as RNN or CNN. However, we use a basic neural network architecture and the MSE as the training objective to investigate the degree of performance improvement provided by GAN.

\begin{figure}[t]
\begin{center}
\vspace{-20pt}
\includegraphics[trim=7.8cm 0.8cm 9cm 1cm,clip,scale=0.42]{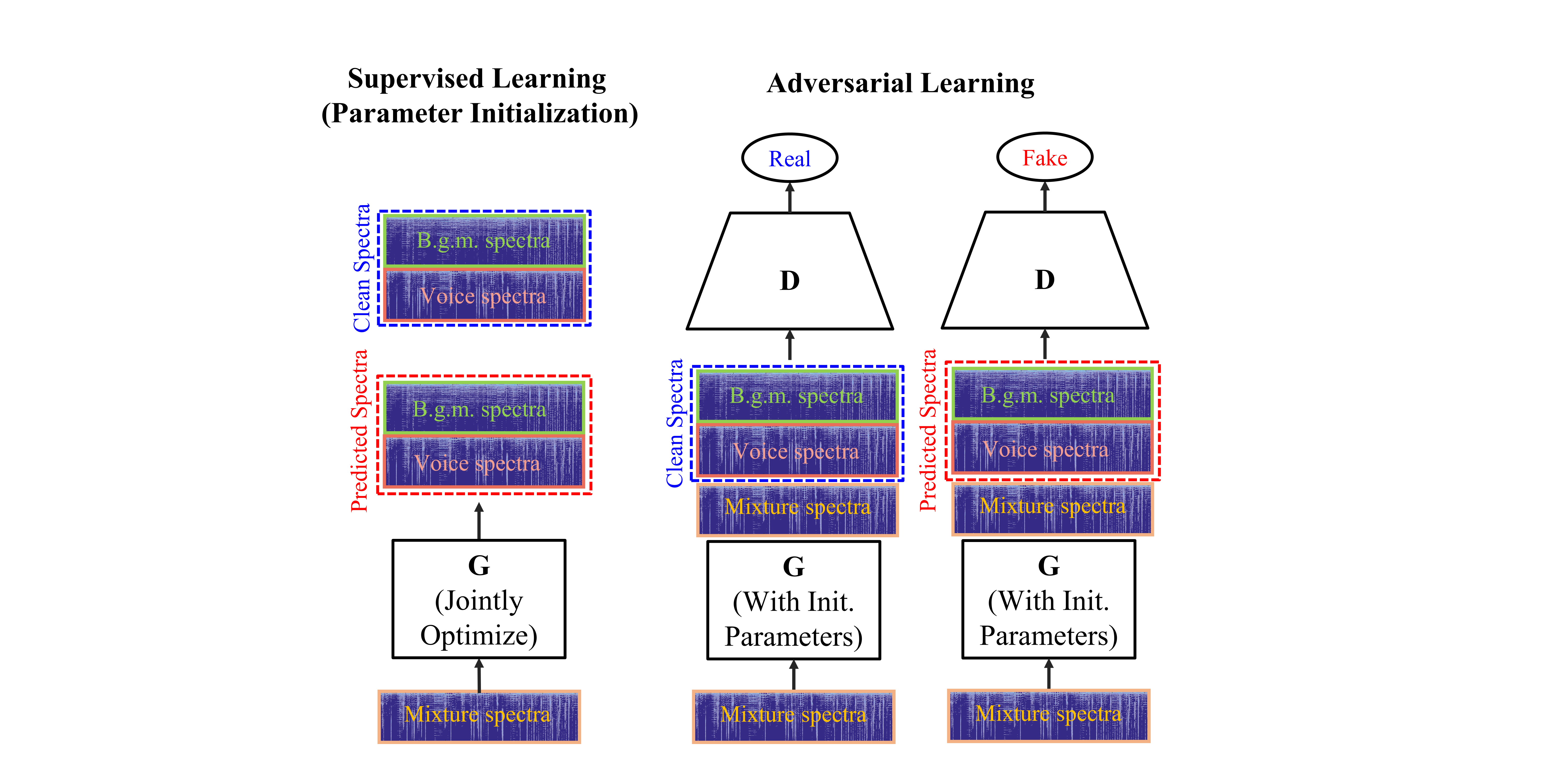}
\caption{SVSGAN training process, where ``B.g.m.'' stands for background music. The parameters of generator G are first inintialized. Discriminator D returns ``fake'' when the input contains predicted spectra and returns ``real'' when input contains clean spectra.}
\label{fig:svsgan_training}
\vspace{-10pt}
\end{center}
\end{figure}

\section{Experiments}
\label{sec:experiments}

\subsection{Dataset \& Settings}
The proposed framework is evaluated using the  MIR-1K dataset\cite{hsu10taslp}, iKala dataset\cite{chan15icassp} and Demixing Secret Database (DSD100) \cite{sisec16mus}. 
The MIR-1K dataset consists of 1,000 song clips lasting 4 to 13 seconds with a sample rate of 16,000 Hz. These clips are recorded from 110 Chinese popular karaoke songs performed by both male and female amateurs. 
The iKala dataset consists of 352 30-second song clips with a sample rate of 44,100 Hz. These clips are recorded from Chinese popular songs performed by professional singers. 
Only 252 song clips are released as a public subset for evaluation.
Each song clip in these two datasets is a stereo recording, with one channel for the singing voice and the other for background music. Manual annotations of the pitch contours are provided.
In experimental settings, we randomly select one-fourth of the song clips for training data and the remaining song clips are used for testing. 

The DSD100 dataset is taken from a subtask called MUS from the Signal Separation Evaluation Campaign (SiSEC). It consists of Dev and Test parts each with 50 songs with a sample rate of 44,100 Hz. Each song provides four sources: bass, drums, other and vocals and the mixture is semi-professionally engineered. The average duration of these songs is 4 minutes and 10 seconds and the dataset includes a wide variety of music genres. 

To reduce computational cost, all song clips from the iKala and DSD100 datasets are downsampled to 22,050 Hz. We used STFT to yield magnitude spectra with a 1024-point window size and a 256-point hop size. Performance is measured in terms of source to distortion ratio (SDR), source to interference ratio (SIR), and source to artifact ratio (SAR), calculated by the Blind Source Separation (BSS) Eval toolbox v3.0 \cite{vincent06taslp}. For the iKala and MIR-1K datasets, overall performance is reported on weighted means of the SDR, SAR and SIR. For the DSD100 dataset, overall performance is reported on median values of SDR based Test part.

\begin{table}[t]
\vspace{-15pt}
\centering  
\begin{tabular}{l|ccc}  
\hline 
\hline
\multicolumn{4}{c}{\bf{MIR-1K Dataset}} \\
\cline{1-4}
Model              &SDR            &SAR            &SIR   \\ \hline 
DNN (baseline)     &6.57           &10.14          &9.84 \\  
SVSGAN (V+B)       &6.69           &10.32          &9.86 \\
SVSGAN (V+M)       &6.73           &10.28          &9.96 \\        
SVSGAN (V+B+M)     &6.78           &10.29          &10.07 \\  
IBM (upper bound)  &13.92          &14.80          &21.96 \\    
\hline
\hline

\multicolumn{4}{c}{\bf{iKala Dataset}} \\
\cline{1-4}
Model              &SDR            &SAR            &SIR   \\ \hline 
DNN (baseline)     &9.74           &11.72          &14.99 \\  
SVSGAN (V+B)       &10.15          &12.48          &14.72 \\
SVSGAN (V+M)       &10.22          &12.78          &14.41 \\        
SVSGAN (V+B+M)     &10.32          &12.87          &14.54 \\  
IBM (upper bound)  &12.30          &14.10          &23.70 \\    
\hline
\hline
\end{tabular}
\caption{Vocal results (in dB) of conventional DNN and SVSGANs on the MIR-1K and iKala datasets. ``IBM'' represents ideal binary mask. Some examples of singing voice separation are provided at \url{http://mirlab.org/demo/svsgan}. }
\label{tab:tab1}
\vspace{-10pt}
\end{table}

\subsection{Experimental Results}

To compare the performance between the conventional DNN and SVSGANs, we construct a conventional DNN, which consists of 3 hidden layers, each with 1024 neurons, denoted as DNN (baseline). The architecture of generator G in the SVSGANs is identical to the baseline and is combined with discriminator D consisting of 3 hidden layers, each with 512 neurons. 
The difference between the SVSGANs is the input spectra of discriminator D, as shown in Table \ref{tab:tab1}, where ``V'' stands for the vocal spectra, ``B'' is the background music spectra, and ``M'' is the mixture spectra. Comparing DNN (baseline) to SVSGANs, the results on iKala and MIR-1K datasets show that SVSGANs enhance performance in terms of SDR and SAR. Comparing different SVSGAN architectures, SVSGAN (V+B) represents the results of the original GAN architecture while SVSGAN (V+M) and SVSGAN (V+B+M) represent the results of the conditional GAN. SVSGAN (V+M) is found to provide better results, indicating that when the input of discriminator D contains the mixture spectra, SVSGAN (V+M) not only learns the mapping from the distribution of mixture spectra to the distribution of clean spectra but also learns a general structure from the mixture spectra at the same time. Comparing SVSGAN (V+M) to SVSGAN (V+B+M), which has more inputs for discriminator, suggests that increasing the number of inputs to discriminator D improves performance.

\begin{figure}[t]
\begin{center}
\vspace{-20pt}
\includegraphics[trim=2.0cm 0.1cm 1cm 0.5cm,clip,scale=0.40]{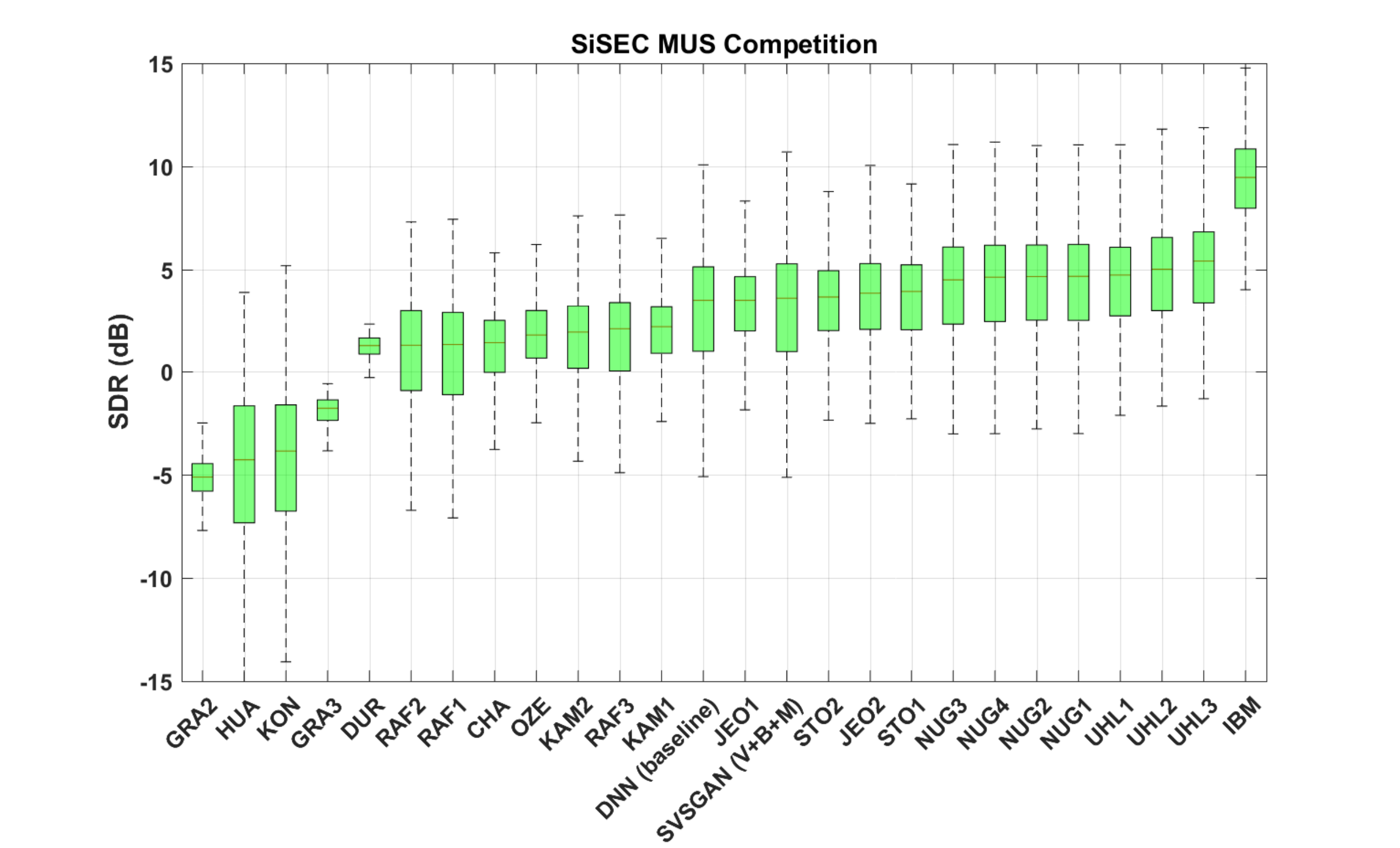}
\caption{Vocal results on the Test part of the DSD100 dataset, sorted by median values of each submission.}
\label{fig:dsd100_result}
\vspace{-10pt}
\end{center}
\end{figure}

Fig.~\ref{fig:dsd100_result} compares the Test part of the DSD100 dataset. The DNN (baseline) and SVSGAN (V+B+M) are the same as those evaluated on the iKala and MIR-1K datasets. Since we only trained the model with the Dev part of dataset without additional augmented datasets, such as MedleyDB\cite{salamon14ismir}, SVSGAN (V+B+M) does not outperform all other submissions. However, the result still shows that singing voice separation can be improved by adversarial learning on this dataset featuring a wide variety of music genres.

\section{Conclusions \& Future work}
\label{sec:concl&future work}

This paper proposes a singing voice separation model with time-frequency masking function for monaural recordings using a generative adversarial framework. The framework consists of two conventional neural networks with conditional GAN, and is shown to potentially enhance source separation performance. Possible future work involves three directions. First, we will incorporate additional augmented data in our adversarial training process to achieve better performance. Next, we will seek to improve generator G and discriminator D using more powerful neural networks, such as CNN or RNN. Finally, we will explore the use of Wasserstein GAN\cite{wgan16arxiv} to achieve better performance. Future work will also include further comparisons between SVSGANs and other competitive approaches.



\vfill\pagebreak

{\small
\bibliographystyle{IEEEbib}
\bibliography{refs}
}
\end{document}